\documentclass[12pt,a4paper]{extarticle}

\usepackage[utf8]{inputenc}
\usepackage[T1,T2A]{fontenc}
\usepackage[russian,english]{babel}

\usepackage[a4paper,top=2.5cm,bottom=2.5cm,left=2.5cm,right=2.5cm,marginparwidth=1.75cm]{geometry}

\usepackage{amssymb,amsmath}
\usepackage{slashed}
\usepackage{mathtools}
\usepackage{feynmf}

\usepackage{mathrsfs}
\usepackage{graphicx}
\usepackage{simpler-wick}

\usepackage[export]{adjustbox}
\usepackage[colorlinks=true, allcolors=blue]{hyperref}
\usepackage{hyperref}
\usepackage{authblk}

\usepackage{tikz}
\usepackage{subfigure}
\usepackage{indentfirst} 
\usepackage{gensymb}

\newcommand{\green}[1]{\textbf{\color{green} [#1]}}

\graphicspath{{Pics/}}

\usepackage{csquotes}
\usepackage{tabu}
\usepackage[style=nature,
sorting=none,
language=autobib,
autolang=other,
doi=false,
url=false
]{biblatex}
\usepackage{url}
\usepackage{aas_macros}

\title{
Hot primordial regions with anomalous hydrogenless chemical composition}

\author[1]{K.M. Belotsky}
\author[1,2]{M. M. El Kasmi}
\author[1,3]{S.G. Rubin}
\author[1]{M.L. Solovyov}
\affil[1]{National Research Nuclear University MEPhI (Moscow Engineering Physics Institute), Kashirskoe shosse 31, 115409 Moscow, Russia.}
\affil[2]{Physics Department, Faculty of Science, Sohag University, Sohag Center, Sohag 82524, Egypt.}
\affil[3]{N.~I.~Lobachevsky Institute of Mathematics and Mechanics, Kazan  Federal  University, Kremlevskaya  Street 18, Kazan 420008, Russia.}
\date{}

\begin{document}

\maketitle

\begin{abstract}
We study primordial nucleosynthesis 
in hypothetical hot regions that could be formed by the primordial density inhomogeneities. It is shown that the regions survived up to the present times acquire an abnormally high metallicity. This conclusion holds in wide range of initial parameters of such regions.
We considered the thermonuclear reaction rates 
and estimated abundances 
of deuterium and helium-3 and -4 inside these areas. It has been established that all baryons tend to form helium-4, which is the thermonuclear link in the chain of formation of heavier elements.
\end{abstract}

\section{Introduction}

We suppose that stable hot regions can be formed in the early Universe. This hypothesis was put forward on the basis of the cosmic X-ray observations and IR background \cite{kashlinsky2019electromagnetic}.
The cluster of primordial black holes (PBH) can 
be responsible for such regions. 
Formation of PBH clusters and their possible observational effects are now of special interest \cite{Rubin:2000dq,Rubin:2001yw,Khlopov:2004sc,Ding:2019tjk,Matsubara:2019qzv,Young:2019gfc,Kawasaki:2021zir,Inman:2019wvr,Afshordi:2003zb,Jedamzik:2020ypm,DeLuca:2020jug,Pilipenko:2022emp},
but we do not constrain possibility of such regions appearance by PBH clusters only.
PBH and their cluster formation can be the consequence of existence and breaking of some new symmetry in quantum field theory \cite{Rubin:2000dq,konoplich1998formation,deng2018cmb}. 

PBH cluster can be the seed of a quasar or galaxy formation \cite{Dokuchaev:2004kr, Dokuchaev:2008hz,Khromykh:2019yyx}. Here we consider the matter trapped in this region, which can be protogalaxy or exist separately. So the prerequisites for the task in question are the regions decoupled from Hubble flow (and virialized) containing primordial plasma. Plasma must flow out to the surroundings by diffusion in the CMB field. If the region is big enough, it can survive to the present time, as it was obtained for the antimatter domain \cite{Khlopov:1998uy, Khlopov:2000as}. 
The region of the size $\sim 1$ pc spreads over ambient matter after recombination ($z\approx 1000$), when the structure forms already (see, e.g., Eq.(12) from \cite{Khlopov:1998uy}). By this time, heavy chemical elements, as we show, have time to be formed in a wide range of considered model parameters, therefore areas contaminated with heavy elements can be expected to exist even if the matter had spread outside primordial region.
Moreover, cooling by conventional thermal (gamma-ray) radiation is ineffective for big regions. The escaping time of photons from the region interior (thermal time scale) at the taken parameters (given below) exceeds the modern age of the Universe. The matter inside the area can be additionally heated with respect to the surrounding one during its formation due to domain wall kinetic energy in the case of the respective mechanism of PBH cluster formation \cite{Rubin:2000dq,Rubin:2001yw,Khlopov:2004sc} including Higgs field \cite{Belotsky:2017puc,Belotsky:2017txw}. 


We consider the chemical composition of such possible hot regions, whatever their origin is. The thermal evolution of such regions involves many factors. 
The matter inside areas can be heated or cooled by various processes acting at the same time. These processes include the neutrino cooling \cite{belotsky2020neutrino,belotsky2020neutrino2}, inelastic reactions between elementary particles and nuclei, the radiation from star-forming hot plasma \cite{kashlinsky2019electromagnetic}, 
the gravitational dynamics of the system, the shock waves and diffusion of matter during the region formation \cite{belotsky2017local, axioms9020071}, energy transfer from collapsing walls mentioned above \cite{konoplich1998formation,deng2018cmb,Dokuchaev:2004kr,berezin1983thin,Davoudiasl:2021ijv}, accretion \cite{kashlinsky2019electromagnetic,axioms9020071} and the Hawking evaporation \cite{axioms9020071, dolgov1993baryon,dolgov2018massive}.
%
We focus here on the pure effect of inelastic reactions between elementary particles and nuclei. They may play a dominant role within a wide range of the region parameters which are specified below. We have shown earlier \cite{belotsky2020neutrino,belotsky2020neutrino2}, that neutrino emission can be decisive in the temperature evolution of such regions at the first stage. Here we extend consideration by involving reactions with the lightest element formation.

We use the results obtained in \cite{belotsky2020neutrino, belotsky2019clusters,Khromykh:2019yyx, Dokuchaev:2004kr, Dokuchaev:2008hz}, where the mass of the detached region was supposed to range $10^4-10^8 M_{\odot}$\footnote{These values have been of interest since they can provide a seed for supermassive black holes and galaxies. We do not relate the amount of PBHs with dark matter, which is strongly constrained in dependence on PBH mass value \cite{Carr:2021bzv}. Abundances and masses of PBHs inside clusters as well as of 
clusters themselves are assumed to be proper ones.}. The following are the most important starting parameters: the area has a radius of $R\sim 1$ pc, a mass of $10^4 \, M_{\odot}$, and an initial temperature interval $T_0\sim 1\text{ keV}- 10$ MeV. 
%
%

The goal of this work is to investigate certain reaction networks, which define the evolution of temperature and chemical composition of the regions in the early Universe. 
Light element abundance ratios ($n_d/n_B$, $n_{^3He}/n_B$ and $n_{^4He}/n_B$) are finally obtained, heavier element production is discussed.

Hypothesis on existence of the regions discussed can be supported by the evidences of cosmic infrared and X-ray background correlations \cite{kashlinsky2019electromagnetic}, anomalous star existence \cite{2018AJ....156..113B, Dolgov:2019ncq}, and can be probed in direct searches for large areas with abnormal chemical composition in future. Also, such sources of high temperature radiation at the pre-recombination stage can give specific observed patterns of CMB temperature variations ($\Delta$T/T) \cite{Grachev:2010tn} because it is determined by the interaction of these fluctuations in matter density with the CMB during the Universe's expansion and cooling, which are not applicable to small scales.

Section \ref{Nucl} is dedicated to the discussion on the main nuclear reactions. Subsection \ref{temp_sec} contains the information about the region temperature, subsection \ref{pn_sec} -- about proton and neutron abundances, \ref{dhe_sec} -- deuterium and helium-3, \ref{hehe_sec} -- about abundances of helium-4 and heavier elements.  A closing overview of the research is provided in section \ref{concl}. We also include some useful information on reaction rates and cross-sections in appendix \ref{reac_app}.

\section{Nucleosynthesis}
\label{Nucl}

Consider the reaction between two nuclei 1 and 2. The  reaction rate is proportional to the mean lifetime $\tau$ of the nuclear species in the stellar plasma. The number density change rate of nucleus 1 caused by reactions with nucleus 2 can
be expressed as
\cite{burbidge1957synthesis,clayton1983principles}
\begin{equation}\label{N122}
\Big(\frac{dn_1}{dt}\Big)_2=-(1+\delta_{12})r_{12}=-(1+\delta_{12})\frac{n_1n_2\langle \sigma v\rangle_{12}}{(1+\delta_{12})}=-n_1n_2\langle \sigma v\rangle_{12}.
\end{equation}
Here $r_{12}$ is the rate of interaction, $\delta_{12}$ is the Kronecker symbol equals one if 1 = 2 and zero if 1 $\neq$ 2, $n_1$ and $n_2$ are the number densities of nuclei of type 1 and type 2 (having the atomic numbers $Z_1$ and $Z_2$, as well as the mass numbers $A_1$ and $A_2$), and $\langle \sigma v\rangle_{12}$ represents the product of the reaction cross section and the interacting nuclei's relative velocity $v$. The case of identical initial nuclei is taken into account by the presence of the Kronecker symbol.

We will look at how the neutrons, protons, $^2$H, $^3$He and $^4$He abundances change over time due to the reactions of mostly proton-proton chain. The n + p and p + p reaction produces $^2$H, which is then destroyed by the d + p and d + $\gamma$ reactions, whereas the d + p reaction produces $^3$He, which is then destroyed by the $^3$He + $^3$He reaction producing $^4$He. We consider neutrinos to be able to leave the region freely and therefore cool it down. The essential reactions of light elements and neutrinos produced are the following:
\begin{align}
e^- + p &\longrightarrow  n + \nu_e, \label{eq_ep} &&\\
e^+ + n &\longrightarrow p + \bar\nu_e, \label{eq_en} &&\\
e^+ + e^- &\longrightarrow \nu_{e,\mu,\tau}+\bar{\nu}_{e,\mu,\tau}, \label{eq_ee} &&\\
n &\longrightarrow p+e^- + \bar\nu_e, \label{n_eq}&&\\
p + p &\longrightarrow D + e^{+} +\nu_{e} + 1.44 \; \text{MeV}, &&\\
D + \gamma &\longrightarrow p + n - 2.22\; \text{MeV}, \label{dg_eq} &&\\
n+ p &\longrightarrow D + \gamma + 2.22 \; \text{MeV}, \label{npd_eq}&&\\
D + p &\longrightarrow \textsuperscript{3}He + \gamma + 5.493 \; \text{MeV}, &&\\
^{3}He + \textsuperscript{3}He &\longrightarrow \textsuperscript{4}He + 2 p + 12.861\; \text{MeV}. \label{He4_eq}
\end{align} 
%
We neglected energy releases of less than 1 MeV.
The initial number densities are approximately described as
\begin{eqnarray}
n_p=\frac{n_B}{1+\exp\left(-\frac{\Delta m}{T_0}\right)},\;\;\;\;\;\;\;\;\;\;\;\;\;\;\;\;\;\;\;\;\;\;\;\;\;\;\; n_n=n_p(T_0)\exp\left(-\frac{\Delta m}{T_0}\right),\label{np_eq}\\
n_{e^-}=n_e^{eq}(T_0)\exp\left(-\frac{m_e}{T_0}\right)+\Delta n_e,\;\;\;\;\;\;\;\;\;\;n_{e^+}= n_e^{eq}(T_0) \exp\left(-\frac{m_e}{T_0}\right),\label{ne}\\
n_B\equiv n_p+n_n=g_B\, \eta n_{\gamma}(T_0),\;\;\;\;\;\;\;\;\;\;\;\;\;\;\;\;\;\; \quad \Delta n_e\equiv n_{e^-}-n_{e^+}=n_p.
\label{nd}
\end{eqnarray}
\\
Here $\eta=n_B/n_{\gamma}\approx 0.6 \times 10^{-9}$ is the baryon to photon relation in the modern universe, $g_B\sim 1$ is the correction factors of that relation due to entropy re-distribution, $n_{\gamma}(T)=\frac{2\zeta(3)}{\pi^2}T^3$ and $n_e^{eq}(T)=\frac{3\zeta(3)}{2\pi^2}T^3$ are the equilibrium photon and electron number densities respectively,  $\Delta m=m_n-m_p=1.2$ MeV. The forms of Equations \eqref{np_eq} and \eqref{ne} for number densities are chosen to fit their asymptotics in the case of thermodynamic equilibrium.

We consider all densities to be independent on space coordinates within the region. The equations \eqref{ne} are also used to calculate electron and positron current number densities with $T$ instead of $T_0$ and total electric charge instead of $n_p$ inside of $\Delta n_e$.

The rates per unit volume, $\gamma_i\equiv \Gamma_i/V$, for reactions listed above are respectively
\begin{eqnarray}
\label{gamma}
\gamma_{ep}=n_{e^-}n_p\langle \sigma v\rangle_{ep},\;\;\;\;\;\;\;\;\;\;\;\;\;\;\;\;\;\;\;\;\;\gamma_{en}=n_{e^+}n_n\langle \sigma v\rangle_{en},\\
\gamma_{ee}=n_{e^-}n_{e^+}\langle \sigma v\rangle_{ee},\;\;\;\;\;\;\;\;\;\;\;\;\;\;\;\;\;\;\;\;\;\;\;\;\;\;\;\;\;\;\;\;\;\;\;\gamma_n=\frac{n_n}{\tau_n},\\
\gamma_{pp}=\frac{n_{p}^2}{2}\langle \sigma v\rangle_{pp},\;\;\;\;\;\;\;\;\;\;\;\;\;\;\;\;\;\;\;\;\;\;\;\;\;\;\;\gamma_{\gamma d}=n_{\gamma}n_{d}\langle \sigma v\rangle_{\gamma d},\\
\gamma_{np}=n_{n}n_p\langle \sigma v\rangle_{np},\;\;\;\;\;\;\;\;\;\;\;\;\;\;\;\;\;\;\;\;\;\;\;\;\gamma_{dp}=n_{d}n_p\langle \sigma v\rangle_{dp}, \\
\gamma_{^3He^3He}=\frac{(n_{^3He})^2}{2}\langle \sigma v\rangle_{^3He^3He}.
\end{eqnarray}
Here $n_i$ is the concentration of the respective species, $\langle \sigma v\rangle_{ij}$  is the reaction rate of interacting particles $i\, {\rm and }\,j$, $v$ is their relative velocity, for reactions \eqref{eq_ep} -- \eqref{eq_ee} $v\simeq 1$ 
and $\tau_n\approx 1000$ s is the neutron lifetime. The electron-electron, electron-proton and electron-neutron cross section are given by Eqs.\eqref{EE} and \eqref{SW} of Appendix.\\

The temperature balance is defined by the first law of thermodynamics
\begin{equation}\label{eq5}
\Delta Q = \delta U,
\end{equation}
where $\Delta Q$ and $\delta U$ are the heat and inner energy gains (in fact, a decrease) of the matter inside the heated area, respectively.
Expanding all the values one obtains
\begin{multline}\label{temp1}
\left[(\gamma_{pp}\cdot Q_1 - \gamma_{\gamma d}\cdot Q_2 + \gamma_{np}\cdot Q_3 + \gamma_{dp}\cdot Q_4 + \gamma_{^3He^3He}\cdot Q_5)-\right. \\ 
\left. ( \gamma_{ep} + \gamma_{en} + 2\gamma_{ee} + \gamma_{n} + \gamma_{pp} )  E_{\nu}\right] dt = 4b T^3 dT,
\end{multline}
where $Q_i$ is energy release of the respective reaction, 
$E_{\nu}\sim T$ is the energy of outgoing neutrino, $b = \pi^2 /15$ is the radiation constant. 
Using Eq. \eqref{N122} and \eqref{temp1} and reactions  \eqref{eq_ee} - \eqref{He4_eq}, we can compose the following system of differential equations.

\begin{align}
\frac{d(n_n)}{dt}&= n_{e^-}n_p\langle \sigma v\rangle_{e^-p}+n_{\gamma}n_{d}\langle \sigma v\rangle_{\gamma d} -\frac{n_n}{\tau_n}-n_{n}n_p\langle \sigma v\rangle_{np}-n_{e^+}n_n\langle \sigma v\rangle_{e^+n} \label{eq_nn} &&\\ 
\frac{d(n_p)}{dt}&= n_{e^+}n_n\langle \sigma v\rangle_{e^+n} + \frac{n_n}{\tau_n} + n_{\gamma}n_{d}\langle \sigma v\rangle_{\gamma d} + (n_{^3He})^2\langle \sigma v\rangle_{^3He^3He} \nonumber \\& - n_{e^-}n_p\langle \sigma v\rangle_{e^-p} - n_{p}^2\langle \sigma v\rangle_{pp} - n_{d}n_p\langle \sigma v\rangle_{dp}  \label{eq_pp}&&\\ 
\frac{d(n_d)}{dt}&= \frac{n_{p}^2}{2}\langle \sigma v\rangle_{pp} + n_{n}n_p\langle \sigma v\rangle_{np} - n_{d}n_{p}\langle \sigma v\rangle_{dp} - n_{\gamma}n_{d}\langle \sigma v\rangle_{\gamma d} \label{eq_DD} &&\\
\frac{d(n_{^3He})}{dt}&=n_{d}n_{p}\langle \sigma v\rangle_{dp}-(n_{^3He})^2\langle \sigma v\rangle_{\text{$^3$He$^3$He}} \label{eq_3He} &&\\ 
\frac{d(n_{^4He})}{dt}&=\frac{(n_{^3He})^2}{2}\langle \sigma v\rangle_{\text{$^3$He$^3$He}} \label{eq_4He}&&\\
\frac{d(T)}{dt}&=[(\gamma_{pp}\cdot Q_1 - \gamma_{\gamma d}\cdot Q_2 + \gamma_{np}\cdot Q_3 + \gamma_{dp}\cdot Q_4 + \gamma_{^3He^3He}\cdot Q_5) \nonumber \\&  -(\gamma_{en} + \gamma_{ep} + 2\gamma_{ee} + \gamma_{n} + \gamma_{pp} )  E_{\nu}]/4b T^3 \label{eq_T}
\end{align}

The initial number densities of deuterium and helium are considered to be zero inside the region.

As can be seen from the equations, we do not consider any reactions of heavy elements production for the sake of simplicity. 
Evidently, some parts of $^4$He will be transformed into heavier elements subsequently, so our estimations of its number density effectively show the number density of $^4$He together with all heavier elements.

\subsection{Temperature evolution}
\label{temp_sec}
The temperature evolution (Eq.\eqref{eq_T}) follows from the equation system above. It is dominated by the cooling due to the reaction \eqref{eq_ee}.  
\begin{figure}[h]
    \centering
    \includegraphics[width=0.9\textwidth]{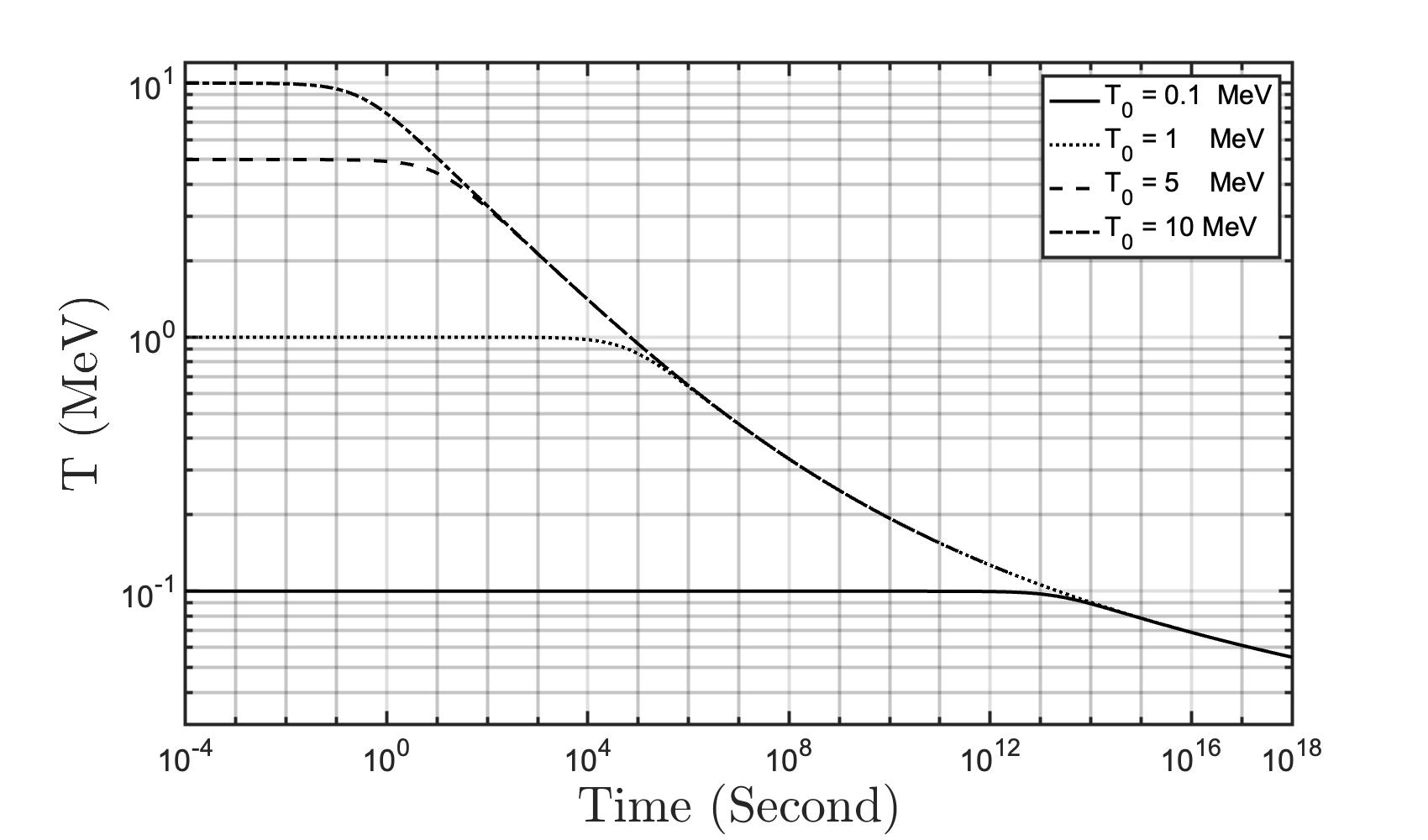}
    \caption{The time behaviour of the temperature inside the heated area.}
    \label{T}
\end{figure}
Figure \ref{T} shows the time dependence of the temperature for different initial temperatures $T_0$. 

\subsection{Abundances of free protons and neutrons}
\label{pn_sec}

\begin{figure}[ht]
    \subfigure{
    \includegraphics[width=0.49\textwidth]{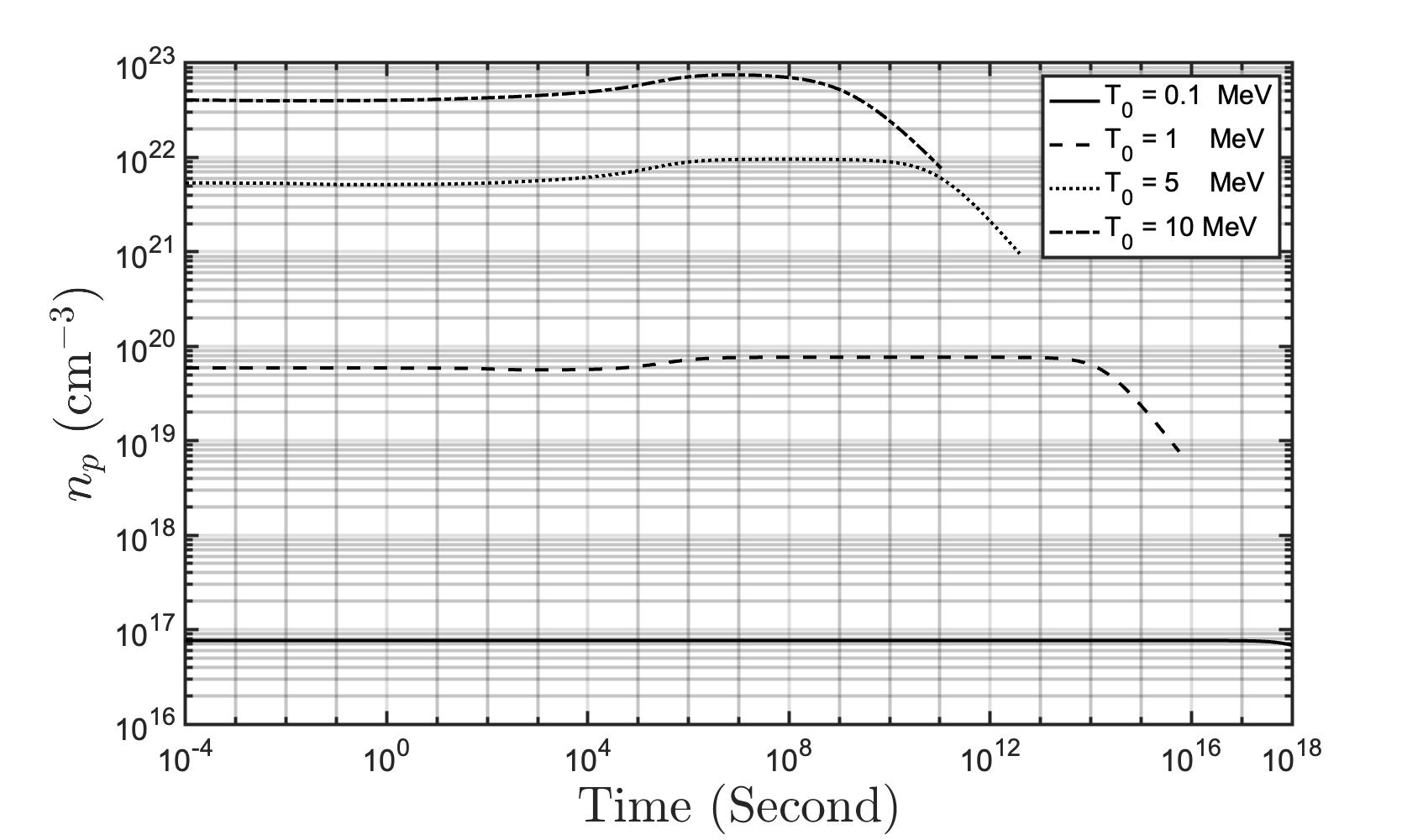}}
    \subfigure{
    \includegraphics[width=0.49\textwidth]{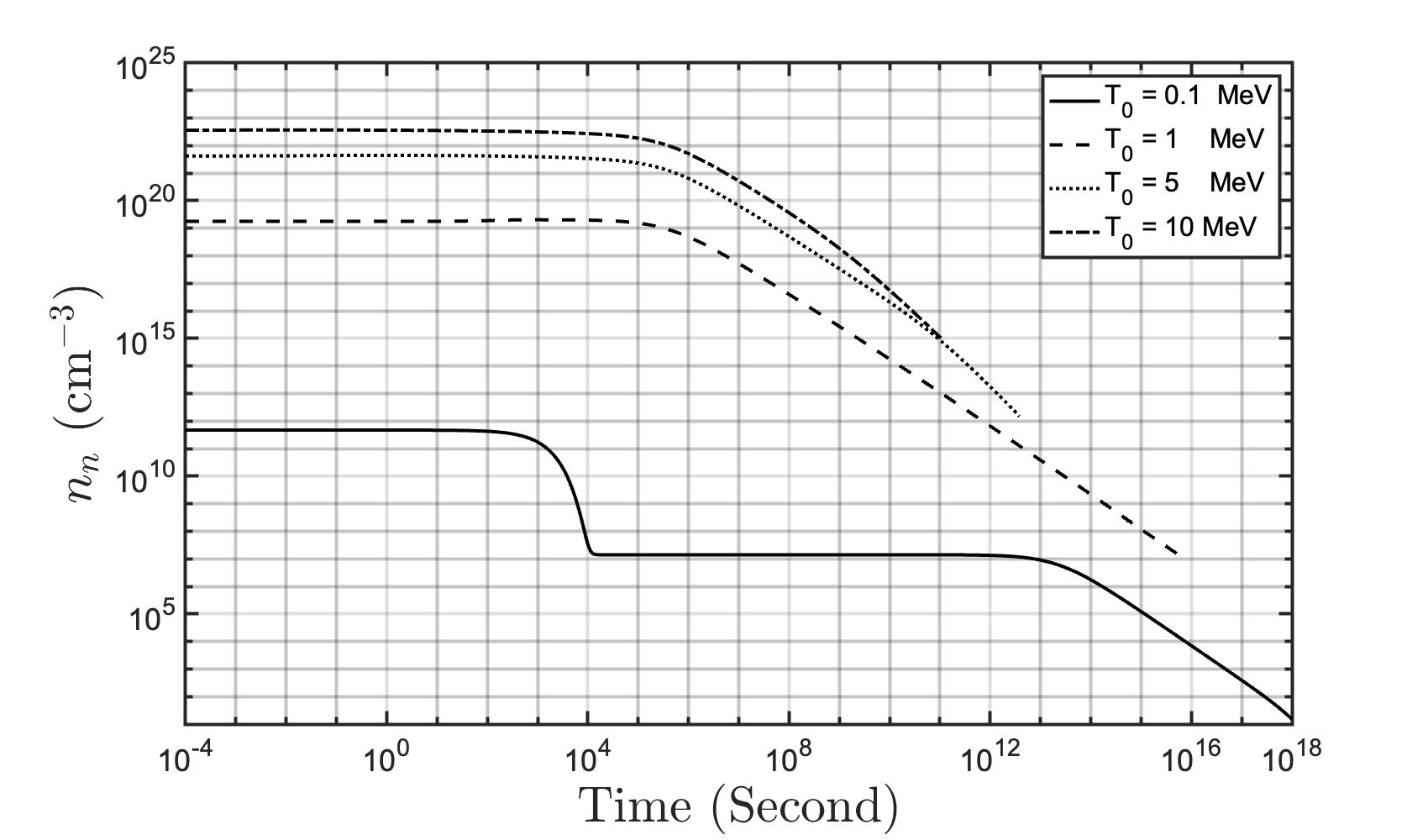}}
    \caption{Left: The time evolution of the protons density in the region at different initial temperatures. Right: The time evolution of the neutrons density in the region at different initial temperatures.} 
    \label{fig001}
\end{figure}
\begin{figure}[ht]
    \subfigure{
    \includegraphics[width=0.49\textwidth]{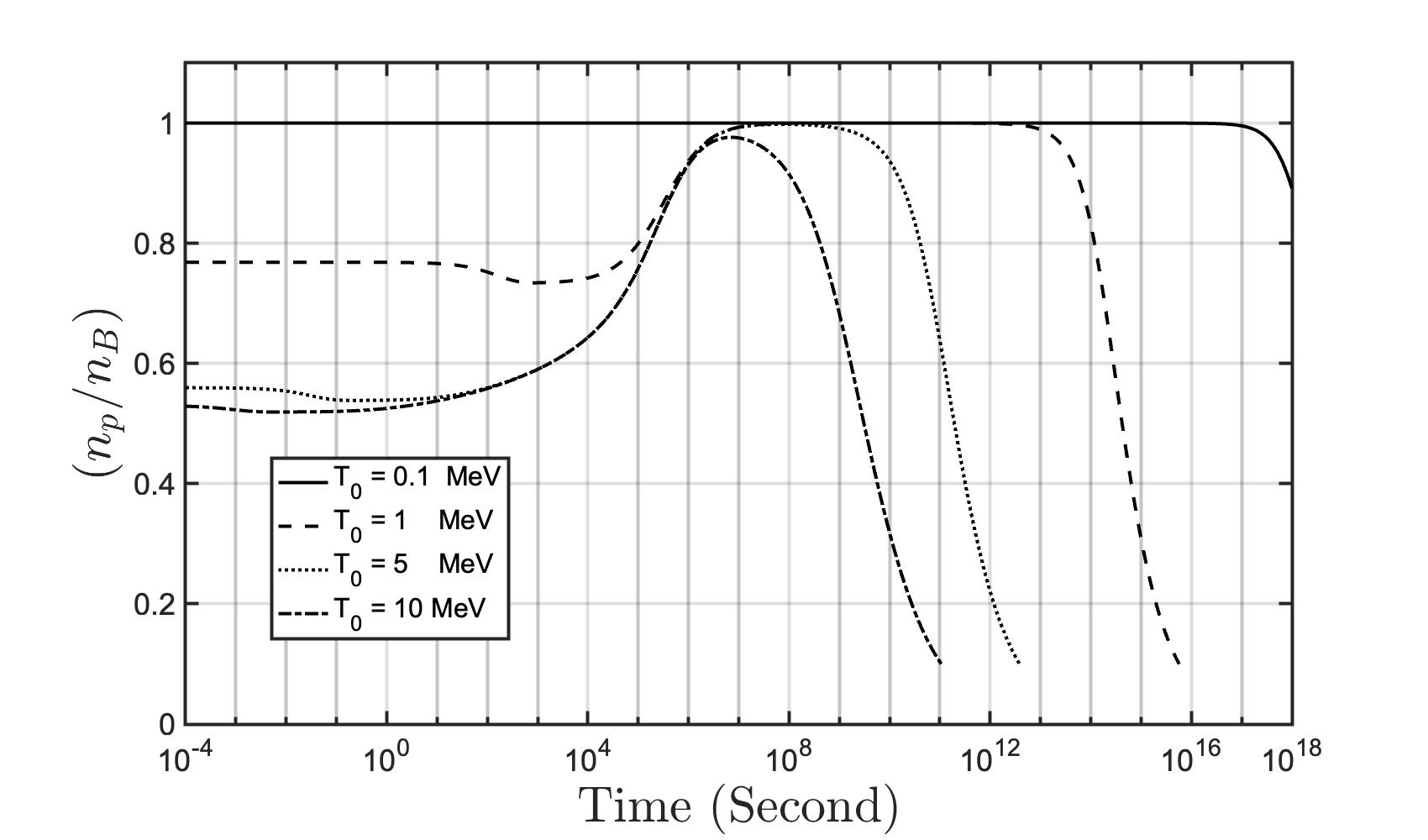}}
    \subfigure{
    \includegraphics[width=0.49\textwidth]{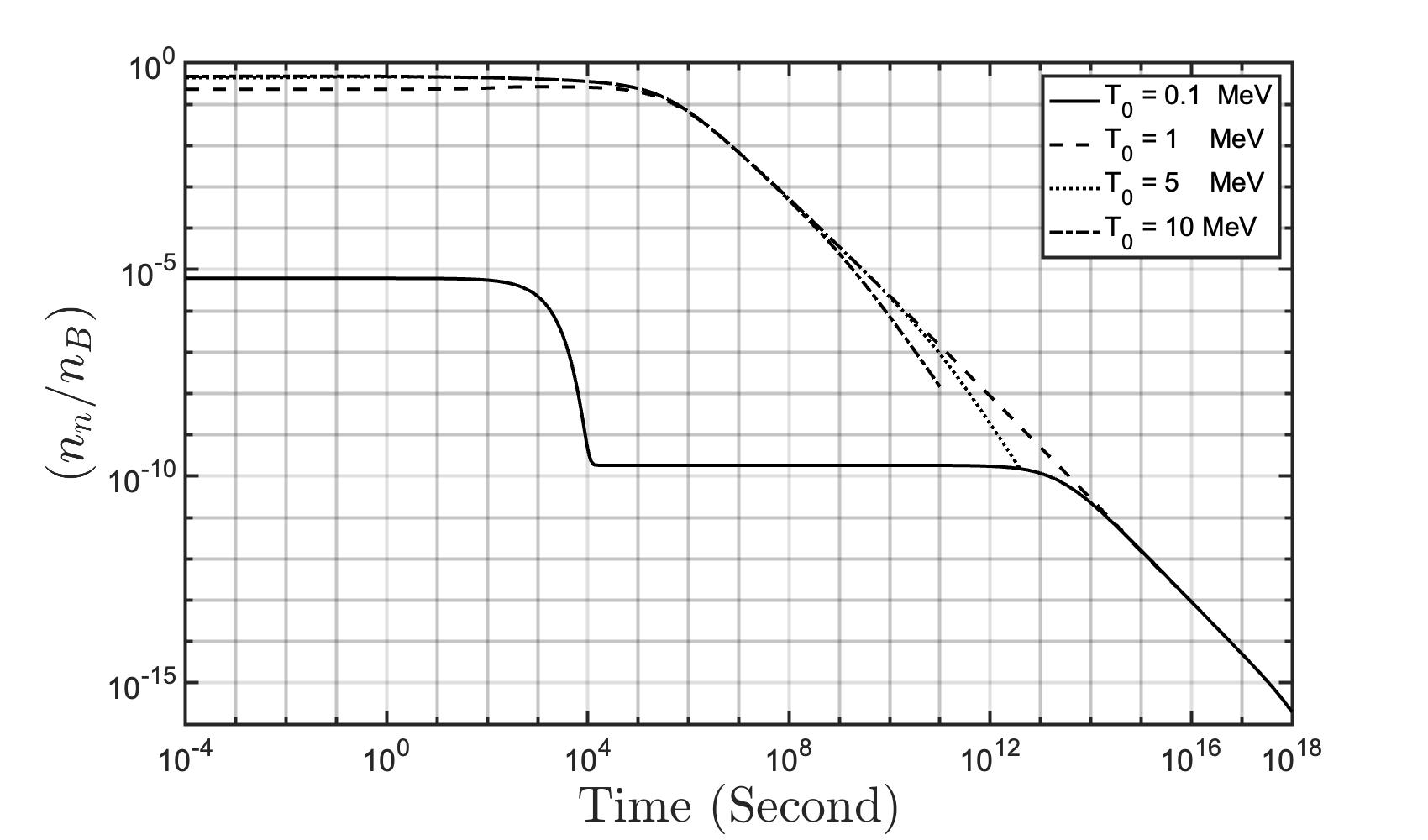}}
    \caption{Left: The time evolution of the abundance protons in the region at different initial temperatures. Right: The time evolution of the abundance neutrons in the region at different initial temperatures.} 
    \label{fig001ex}
\end{figure}

We can estimate the abundance of (free) neutrons and protons numerically using Eqs.\eqref{eq_nn} and \eqref{eq_pp}. Figure \ref{fig001} shows the evolution of the number densities, while Figure \ref{fig001ex} shows the fraction of protons (left) or neutrons (right) from the initial baryon number density.

One can explain qualitatively these figures. There are five processes considered to affect the neutron number density. However, while the production (Eq. \eqref{npd_eq}) and destruction (Eq.\eqref{dg_eq}) of the deuterium are generally the two most active of them, their reaction rates have almost negligible difference in the most cases. Therefore, the neutron abundance is defined by neutron decays (Eq.\eqref{n_eq}) with the combined effect of electron-proton and positron-neutron reactions (Eqs.\eqref{eq_ep} and \eqref{eq_en}). At the higher initial temperatures the latter starts as dominant, slowly decreasing its effect with the fall of the temperature, until it reaches the level of the neutron decays somewhere below 1 MeV. After that, the combination of all three of these reactions causes the slow and gradual fall of neutron abundance. At low initial temperature neutron decays start as dominant process, causing the exponential drop at around $10^3$ seconds, until the decay rate matches the one of the e-p and e-n combination. After that, the neutron abundance remains stable for a long time until the temperature starts having noticeable changes, affecting the reaction rates and causing the neutron number density to have a slow and gradual fall, as in the case of high initial temperatures.

The rise in proton number density is caused by neutron decays (slowed down due to the processes described above). This effect is more visible at the high initial temperatures, as the neutrons constitute a higher part of the baryons there. The consequent fall in the proton number density is caused by the irreversible transition of the baryons to the $^3$He and $^4$He.

\subsection{Abundances of deuterium and helium-3}
\label{dhe_sec}

\begin{figure}[ht]
    \subfigure{
    \includegraphics[width=0.49\textwidth]{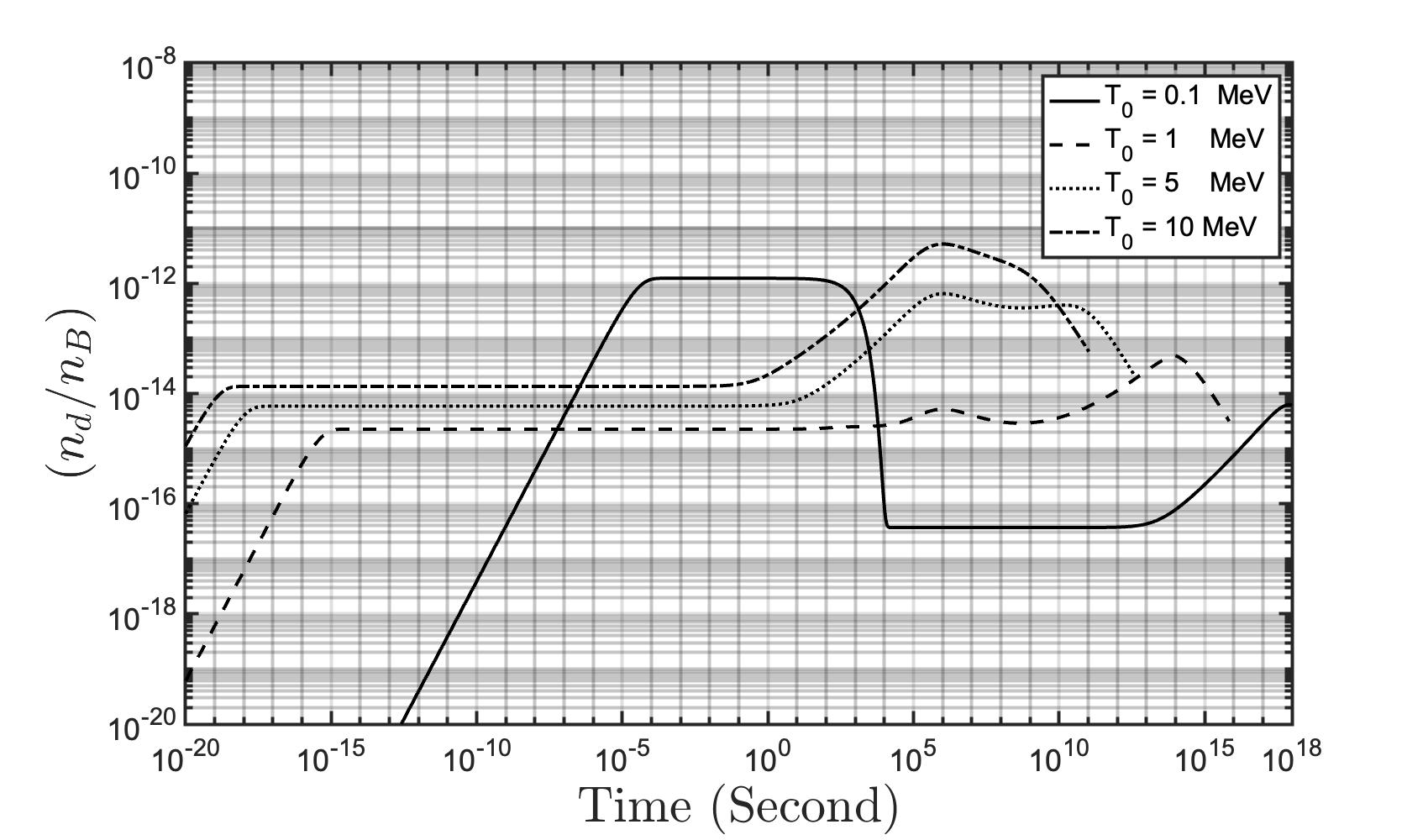}}
    \subfigure{
    \includegraphics[width=0.49\textwidth]{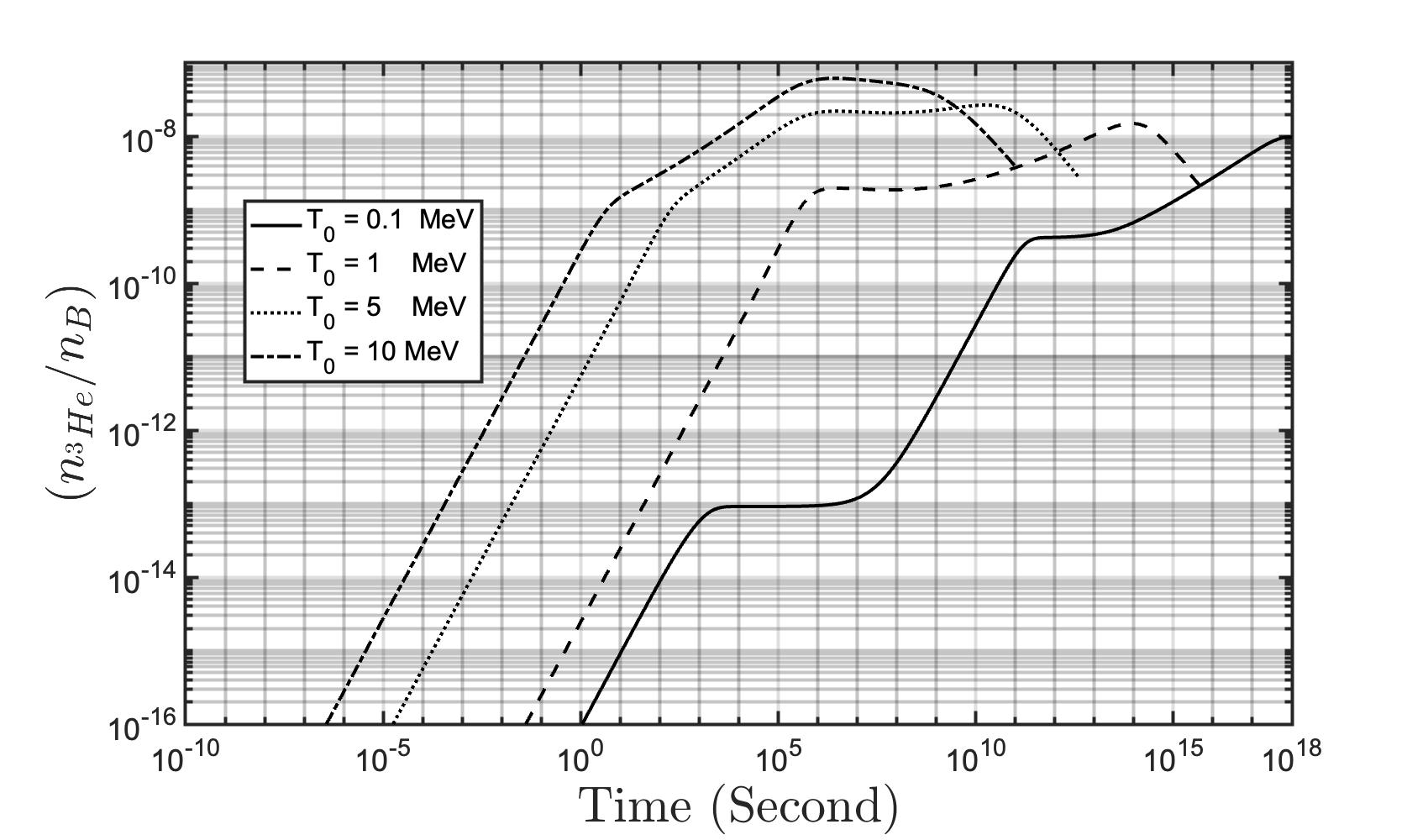}}
    \caption{Left: The time evolution of the abundance ratio $(n_d/n_B)$ in the region at different initial temperatures. Right: The time evolution of the abundance ratio $(n_{^3He}/n_B)$ in the region at different initial temperatures. } 
    \label{fig002}
\end{figure}

Figure \ref{fig002} shows the evolution of deuterium (left panel) and $^3$He fractions with time.

The reaction rates of the deuterium production and destruction equalize themselves under the current values of the temperature and number densities of neutrons and protons in a very short amount of time, reaching the <<equilibrium>>. This equilibrium keeps adjusting to the changes in those values with time.


The abundance of helium-3 is growing most of the time, as the rate of its production is greater than the rate of its destruction into $^4$He. For the high initial temperatures, at late time, this situation reverses due to the decrease in  temperature and number densities of protons and, subsequently, deuterium, and $^3$He starts falling.

The Figure 
\ref{fig002} shows that 
deuterium and helium-3 have very low abundances, making them very likely undetectable. Nonetheless, they play a significant role in the synthesis of heavier elements due to their high reaction rates.

\subsection{Abundance of helium-4 with heavier elements}

\label{hehe_sec}

\begin{figure}[ht]
    \subfigure{
    \includegraphics[width=0.49\textwidth]{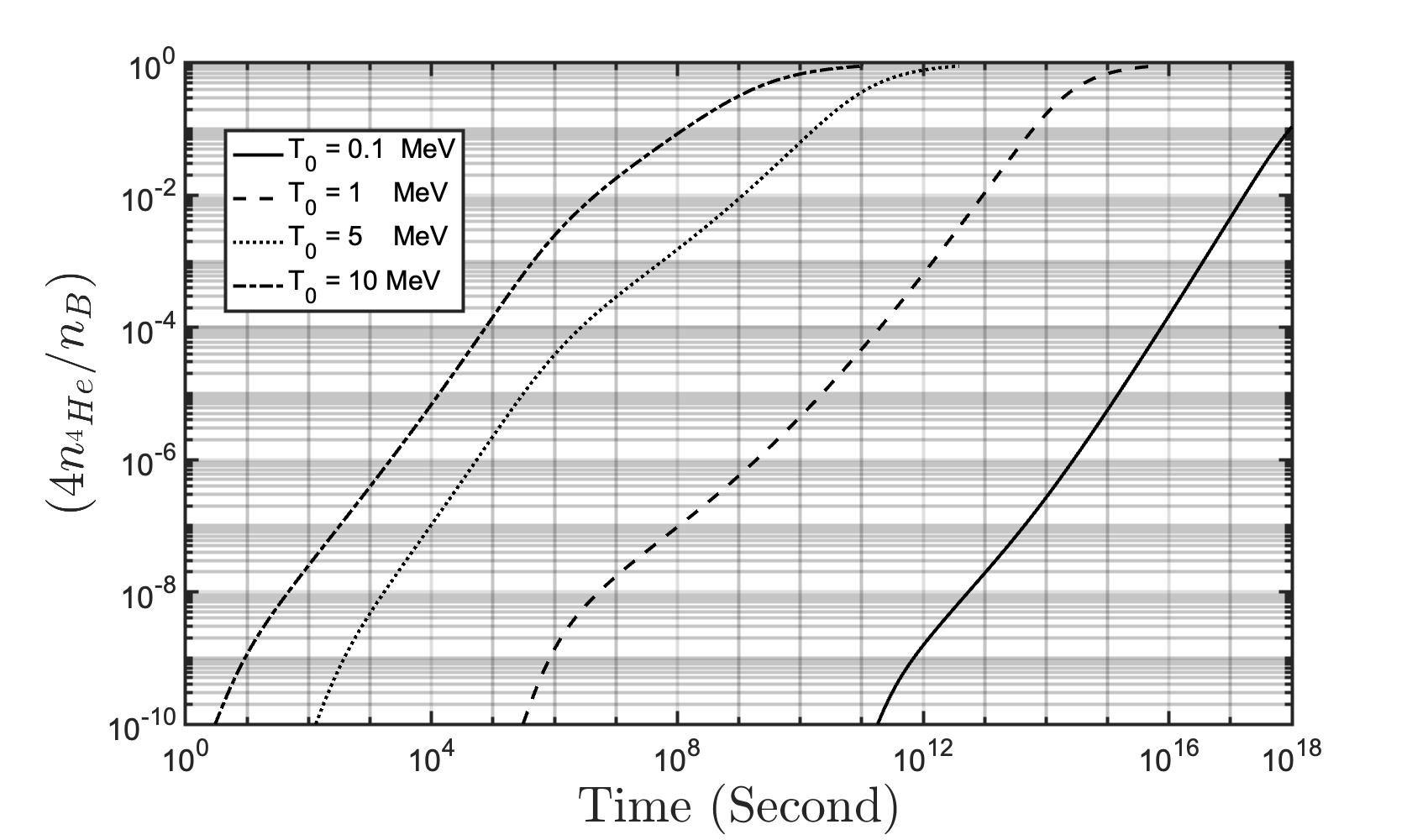}}
    \subfigure{
    \includegraphics[width=0.49\textwidth]{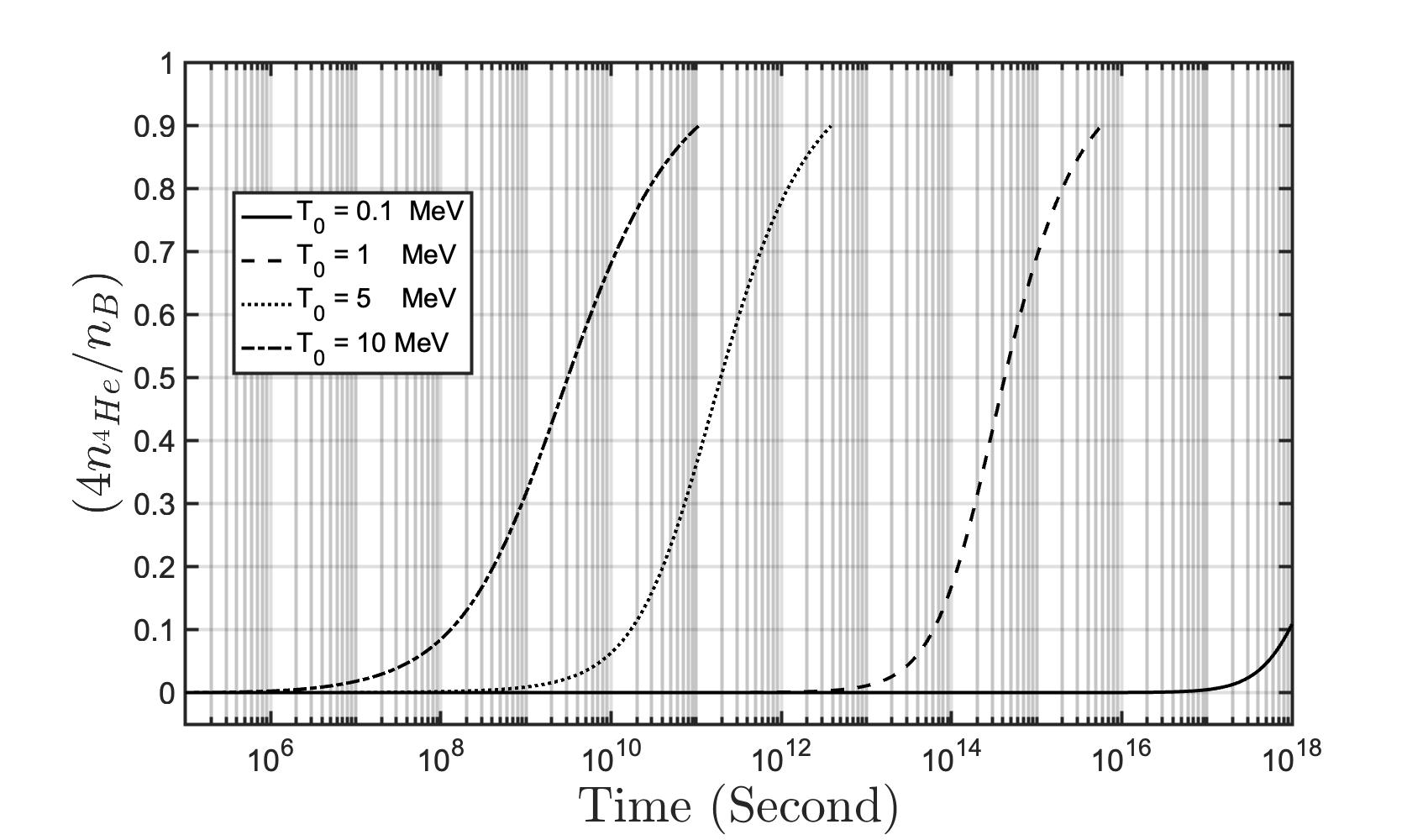}}
    \caption{The time evolution of the density ratio $\rho_{^4He}/\rho_B = 4n_{^4He}/n_B$ in the region at different initial temperatures. Left panel: log-log scale. Right panel: log-linear scale.} 
    \label{fig003}
\end{figure}

We can estimate the helium abundance $(n_{^4He}/n_B)$ numerically from Eq. \eqref{eq_4He}. Figure \ref{fig003} shows the obtained results.

As already stated above, $^4$He abundance here effectively stands for not only helium-4 itself, but also for heavier elements. While our assumptions do not allow us to estimate the metallicity of such a region, we can still make an interesting conclusion that for most of the considered initial temperatures, the dominant part of baryons will be transformed into helium-4 and subsequent elements, leaving the region with almost no hydrogen.

\section{Conclusion}
\label{concl}

We considered the possible existence of stable hot areas 
formed in the early Universe. Their origin could be related to the formation of PBH clusters. There are many factors that affect the evolution of such regions, we focus here on the nuclear reactions inside them. The neutrinos produced in these processes carry away energy, what is found to play a decisive role in temperature change under our approximation (considering nuclear reactions only with the given density and reaction rate dependencies). The considered nuclear reactions tend to form heavy elements, depleting the hydrogen content. The absence of hydrogen in such areas may be a distinguishing feature for their possible search. It will be possible to relate the observed chemical composition to its initial temperature and can account for the existence of anomalous stars.

\textit{Acknowledgment}

The work was supported by the MEPhI Program Priority 2030. 


\printbibliography[heading=bibintoc,title={References}]

@article{kashlinsky2019electromagnetic,
  title={Electromagnetic probes of primordial black holes as dark matter},
  author={Kashlinsky, Alexander and Ali-Ha{\"\i}moud, Y and Clesse, S and Garcia-Bellido, J and Wyrzykowski, L and Achucarro, A and Amendola, L and Annis, J and Arbey, A and Arendt, RG and others},
  journal={Bull. Am. Astron. Soc.},
  year={2019},
  publisher={Elsevier}
}

@article{belotsky2019clusters,
  title={Clusters of primordial black holes},
  author={Belotsky, Konstantin M and Dokuchaev, Vyacheslav I and Eroshenko, Yury N and Esipova, Ekaterina A and Khlopov, Maxim Yu and Khromykh, Leonid A and Kirillov, Alexander A and Nikulin, Valeriy V and Rubin, Sergey G and Svadkovsky, Igor V},
  journal={The European Physical Journal C},
  volume={79},
  number={3},
  pages={1--20},
  year={2019},
  publisher={Springer}
}

@article{belotsky2020neutrino,
  title={Neutrino Cooling of Primordial Hot Regions},
  author={Belotsky, Konstantin and El Kasmi, Mohamed and Rubin, Sergey},
  journal={Symmetry},
  volume={12},
  number={9},
  pages={1442},
  year={2020},
  publisher={Multidisciplinary Digital Publishing Institute}
}

@article{belotsky2020neutrino2,
  title={Neutrino cooling effect of primordial hot areas in dependence on its size},
  author={Belotsky, KM and El Kasmi, MM and Rubin, SG},
  journal={Proceedings to the 23rd Workshop What Comes Beyond the Standard Models Bled. arXiv preprint arXiv:2011.14221
  },
  year={2020}
}

@article{burbidge1957synthesis,
  title={Synthesis of the elements in stars},
  author={Burbidge, E Margaret and Burbidge, Geoffrey Ronald and Fowler, William A and Hoyle, Fred},
  journal={Reviews of modern physics},
  volume={29},
  number={4},
  pages={547},
  year={1957},
  publisher={APS}
}

@book{clayton1983principles,
  title={Principles of stellar evolution and nucleosynthesis},
  author={Clayton, Donald D},
  year={1983},
  publisher={University of Chicago press}
}

@book{iliadis2015nuclear,
  title={Nuclear physics of stars},
  author={Iliadis, Christian},
  year={2015},
  publisher={John Wiley \& Sons: Hoboken, NJ, USA}
}

@article{angulo1999compilation,
  title={A compilation of charged-particle induced thermonuclear reaction rates},
  author={Angulo, Carmen and Arnould, Marcel and Rayet, Marc and Descouvemont, Pierre and Baye, D and Leclercq-Willain, C and Coc, Alain and Barhoumi, Slimane and Aguer, P and Rolfs, Claus and others},
  journal={Nuclear Physics A},
  volume={656},
  number={1},
  pages={3--183},
  year={1999},
  publisher={Elsevier}
}

@article{fowler1967thermonuclear,
  title={Thermonuclear reaction rates},
  author={Fowler, William A and Caughlan, Georgeanne R and Zimmerman, Barbara A},
  journal={Annual Review of Astronomy and Astrophysics},
  volume={5},
  number={1},
  pages={525--570},
  year={1967},
  publisher={Annual Reviews 4139 El Camino Way, PO Box 10139, Palo Alto, CA 94303-0139, USA}
}

@book{lang2013astrophysical,
  title={Astrophysical Formulae: Space, time, matter and cosmology},
  author={Lang, Kenneth R},
  year={2013},
  publisher={Springer: Berlin/Heidelberg, Germany}
}

@article{belotsky2017local,
  title={Local heating of matter in the early universe owing to the interaction of the Higgs field with a scalar field},
  author={Belotsky, KM and Golikova, Yu A and Rubin, SG},
  journal={Physics of Atomic Nuclei},
  volume={80},
  number={4},
  pages={718--720},
  year={2017},
  publisher={Springer}
}

@article{berezin1983thin,
  title={Thin-wall vacuum domain evolution},
  author={Berezin, VA and Kuzmin, VA and Tkachev, II},
  journal={Physics Letters B},
  volume={120},
  number={1-3},
  pages={91--96},
  year={1983},
  publisher={Elsevier}
}

@article{konoplich1998formation,
  title={Formation of black holes in first-order phase transitions in the Universe},
  author={Konoplich, RV and Rubin, SG and Sakharov, AS and Khlopov, M Yu},
  journal={Astronomy Letters},
  volume={24},
  pages={413--417},
  year={1998}
}

@article{deng2018cmb,
  title={CMB spectral distortions from black holes formed by vacuum bubbles},
  author={Deng, Heling and Vilenkin, Alexander and Yamada, Masaki},
  journal={Journal of Cosmology and Astroparticle Physics},
  volume={2018},
  number={07},
  pages={059},
  year={2018},
  publisher={IOP Publishing}
}

@article{dolgov1993baryon,
  title={Baryon isocurvature fluctuations at small scales and baryonic dark matter},
  author={Dolgov, Alexandre and Silk, Joseph},
  journal={Physical Review D},
  volume={47},
  number={10},
  pages={4244},
  year={1993},
  publisher={APS}
}

@article{dolgov2018massive,
  title={Massive and supermassive black holes in the contemporary and early Universe and problems in cosmology and astrophysics},
  author={Dolgov, Aleksandr Dmitrievich},
  journal={Physics-Uspekhi},
  volume={61},
  number={2},
  pages={115},
  year={2018},
  publisher={IOP Publishing}
}

@Article{axioms9020071,
AUTHOR = {Khlopov, Maxim and Paik, Biplab and Ray, Saibal},
TITLE = {Revisiting Primordial Black Hole Evolution},
JOURNAL = {Axioms},
VOLUME = {9},
YEAR = {2020},
NUMBER = {2},
ARTICLE-NUMBER = {71},
URL = {https://www.mdpi.com/2075-1680/9/2/71},
ISSN = {2075-1680},
DOI = {10.3390/axioms9020071}
}

@article{Dokuchaev:2008hz,
    author = "Dokuchaev, V. I. and Eroshenko, Yu. N. and Rubin, S. G.",
    title = "{Early formation of galaxies initiated by clusters of primordial black holes}",
    eprint = "0801.0885",
    archivePrefix = "arXiv",
    primaryClass = "astro-ph",
    doi = "10.1134/S1063772908100016",
    journal = "Astron. Rep.",
    volume = "52",
    pages = "779--789",
    year = "2008"
}

@article{Dokuchaev:2004kr,
    author = "Dokuchaev, Vyacheslav and Eroshenko, Yury and Rubin, Sergei",
    editor = "Alimi, J. M. and Khlopov, M. Yu. and Melnikov, V. N.",
    title = "{Quasars formation around clusters of primordial black holes}",
    eprint = "astro-ph/0412418",
    archivePrefix = "arXiv",
    journal = "Grav. Cosmol.",
    volume = "11",
    pages = "99--104",
    year = "2005"
}

@article{Rubin:2000dq,
    author = "Rubin, S. G. and Khlopov, M. Yu. and Sakharov, A. S.",
    %editor = "Khlopov, M. Yu. and Prokhorov, M. E. and Starobinsky, A. A.",
    title = "{Primordial black holes from nonequilibrium second order phase transition}",
    eprint = "hep-ph/0005271",
    archivePrefix = "arXiv",
    journal = "Grav. Cosmol.",
    volume = "6",
    pages = "51--58",
    year = "2000"
}

@article{Rubin:2001yw,
    author = "Rubin, Sergey G. and Sakharov, Alexander S. and Khlopov, Maxim Yu.",
    title = "{The Formation of primary galactic nuclei during phase transitions in the early universe}",
    eprint = "hep-ph/0106187",
    archivePrefix = "arXiv",
    doi = "10.1134/1.1385631",
    journal = "J. Exp. Theor. Phys.",
    volume = "91",
    pages = "921--929",
    year = "2001"
}

@article{Khlopov:2004sc,
    author = "Khlopov, Maxim. Yu. and Rubin, Sergey G. and Sakharov, Alexander S.",
    title = "{Primordial structure of massive black hole clusters}",
    eprint = "astro-ph/0401532",
    archivePrefix = "arXiv",
    reportNumber = "CERN-PH-TH-2004-004",
    doi = "10.1016/j.astropartphys.2004.12.002",
    journal = "Astropart. Phys.",
    volume = "23",
    pages = "265",
    year = "2005"
}

@article{Ding:2019tjk,
    author = "Ding, Qianhang and Nakama, Tomohiro and Silk, Joseph and Wang, Yi",
    title = "{Detectability of Gravitational Waves from the Coalescence of Massive Primordial Black Holes with Initial Clustering}",
    eprint = "1903.07337",
    archivePrefix = "arXiv",
    primaryClass = "astro-ph.CO",
    doi = "10.1103/PhysRevD.100.103003",
    journal = "Phys. Rev. D",
    volume = "100",
    number = "10",
    pages = "103003",
    year = "2019"
}

@article{Matsubara:2019qzv,
    author = "Matsubara, Takahiko and Terada, Takahiro and Kohri, Kazunori and Yokoyama, Shuichiro",
    title = "{Clustering of primordial black holes formed in a matter-dominated epoch}",
    eprint = "1909.04053",
    archivePrefix = "arXiv",
    primaryClass = "astro-ph.CO",
    reportNumber = "KEK-TH-2155; KEK-Cosmo-243; IPMU19-0128, KEK-TH-2154; KEK-Cosmo-243; IPMU19-0128, KEK-TH-2154; KEK-Cosmo-243",
    doi = "10.1103/PhysRevD.100.123544",
    journal = "Phys. Rev. D",
    volume = "100",
    number = "12",
    pages = "123544",
    year = "2019"
}

@article{Young:2019gfc,
    author = "Young, Sam and Byrnes, Christian T.",
    title = "{Initial clustering and the primordial black hole merger rate}",
    eprint = "1910.06077",
    archivePrefix = "arXiv",
    primaryClass = "astro-ph.CO",
    doi = "10.1088/1475-7516/2020/03/004",
    journal = "JCAP",
    volume = "03",
    pages = "004",
    year = "2020"
}

@article{Kawasaki:2021zir,
    author = "Kawasaki, Masahiro and Murai, Kai and Nakatsuka, Hiromasa",
    title = "{Strong clustering of primordial black holes from Affleck-Dine mechanism}",
    eprint = "2107.03580",
    archivePrefix = "arXiv",
    primaryClass = "astro-ph.CO",
    doi = "10.1088/1475-7516/2021/10/025",
    journal = "JCAP",
    volume = "10",
    pages = "025",
    year = "2021"
}

@article{Inman:2019wvr,
    author = {Inman, Derek and Ali-Ha{\"i}moud, Yacine},
    title = "{Early structure formation in primordial black hole cosmologies}",
    eprint = "1907.08129",
    archivePrefix = "arXiv",
    primaryClass = "astro-ph.CO",
    doi = "10.1103/PhysRevD.100.083528",
    journal = "Phys. Rev. D",
    volume = "100",
    number = "8",
    pages = "083528",
    year = "2019"
}

@article{Afshordi:2003zb,
    author = "Afshordi, N. and McDonald, P. and Spergel, D. N.",
    title = "{Primordial black holes as dark matter: The Power spectrum and evaporation of early structures}",
    eprint = "astro-ph/0302035",
    archivePrefix = "arXiv",
    doi = "10.1086/378763",
    journal = "Astrophys. J. Lett.",
    volume = "594",
    pages = "L71--L74",
    year = "2003"
}

@article{Jedamzik:2020ypm,
    author = "Jedamzik, Karsten",
    title = "{Primordial Black Hole Dark Matter and the LIGO/Virgo observations}",
    eprint = "2006.11172",
    archivePrefix = "arXiv",
    primaryClass = "astro-ph.CO",
    doi = "10.1088/1475-7516/2020/09/022",
    journal = "JCAP",
    volume = "09",
    pages = "022",
    year = "2020"
}

@article{DeLuca:2020jug,
    author = "De Luca, V. and Desjacques, V. and Franciolini, G. and Riotto, A.",
    title = "{The clustering evolution of primordial black holes}",
    eprint = "2009.04731",
    archivePrefix = "arXiv",
    primaryClass = "astro-ph.CO",
    doi = "10.1088/1475-7516/2020/11/028",
    journal = "JCAP",
    volume = "11",
    pages = "028",
    year = "2020"
}

@article{Pilipenko:2022emp,
    author = "Pilipenko, Sergey and Tkachev, Maxim and Ivanov, Pavel",
    title = "{Evolution of a primordial binary black hole due to interaction with cold dark matter and the formation rate of gravitational wave events}",
    eprint = "2205.10792",
    archivePrefix = "arXiv",
    primaryClass = "astro-ph.CO",
    doi = "10.1103/PhysRevD.105.123504",
    journal = "Phys. Rev. D",
    volume = "105",
    number = "12",
    pages = "123504",
    year = "2022"
}

@article{Khlopov:2000as,
    author = "Khlopov, Maxim Yu. and Rubin, Sergei G. and Sakharov, Alexander S.",
    title = "{Possible origin of antimatter regions in the baryon dominated universe}",
    eprint = "hep-ph/0003285",
    archivePrefix = "arXiv",
    doi = "10.1103/PhysRevD.62.083505",
    journal = "Phys. Rev. D",
    volume = "62",
    pages = "083505",
    year = "2000"
}

@article{Khlopov:1998uy,
    author = "Khlopov, M. Yu. and Konoplich, R. V. and Mignani, R. and Rubin, S. G. and Sakharov, A. S.",
    title = "{Physical origin, evolution and observational signature of diffused antiworld}",
    eprint = "astro-ph/9810228",
    archivePrefix = "arXiv",
    reportNumber = "INFN-1228-98",
    doi = "10.1016/S0927-6505(99)00099-7",
    journal = "Astropart. Phys.",
    volume = "12",
    pages = "367--372",
    year = "2000"
}

@article{Khromykh:2019yyx,
    author = "Khromykh, Leonid A. and Kirillov, Alexander A.",
    title = "{The gravitational dynamics of the primordial black holes cluster}",
    doi = "10.1088/1742-6596/1390/1/012090",
    journal = "J. Phys. Conf. Ser.",
    volume = "1390",
    number = "1",
    pages = "012090",
    year = "2019"
}

@ARTICLE{2018AJ....156..113B,
       author = {{Bennett}, David P. and {Udalski}, Andrzej and {Bond}, Ian A. and {Suzuki}, Daisuke and {Ryu}, Yoon-Hyun and {Abe}, Fumio and {Barry}, Richard K. and {Bhattacharya}, Aparna and {Donachie}, Martin and {Fukui}, Akihiko and {Hirao}, Yuki and {Kawasaki}, Kohei and {Kondo}, Iona and {Koshimoto}, Naoki and {Li}, Man Cheung Alex and {Matsubara}, Yutaka and {Miyazaki}, Shota and {Muraki}, Yasushi and {Nagakane}, Masayuki and {Ohnishi}, Koji and {Ranc}, Cl{\'e}ment and {Rattenbury}, Nicholas J. and {Suematsu}, Haruno and {Sumi}, Takahiro and {Tristram}, Paul J. and {Yonehara}, Atsunori and {MOA Collaboration} and {Szyma{\'n}ski}, Micha{\l} K. and {Soszy{\'n}ski}, Igor and {Wyrzykowski}, {\L}ukasz and {Ulaczyk}, Krzysztof and {Poleski}, Radek and {Koz{\l}owski}, Szymon and {Pietrukowicz}, Pawe{\l} and {Skowron}, Jan and {OGLE Collaboration} and {Shvartzvald}, Yossi and {Maoz}, Dan and {Kaspi}, Shai and {Friedmann}, Matan and {Wise Group} and {Batista}, Virginie and {DePoy}, Darren and {Dong}, Subo and {Gaudi}, B. Scott and {Gould}, Andrew and {Han}, Cheongho and {Pogge}, Richard W. and {Tan}, Thiam-Guan and {Yee}, Jennifer C. and {{\ensuremath{\mu}}FUN Collaboration}},
        title = "{A Planetary Microlensing Event with an Unusually Red Source Star: MOA-2011-BLG-291}",
      journal = {\aj},
         year = 2018,
        month = sep,
       volume = {156},
       number = {3},
          eid = {113},
        pages = {113},
          doi = {10.3847/1538-3881/aad59c},
archivePrefix = {arXiv},
       eprint = {1806.06106},
 primaryClass = {astro-ph.EP},
       adsurl = {https://ui.adsabs.harvard.edu/abs/2018AJ....156..113B}
}

@article{Dolgov:2019ncq,
    author = "Dolgov, A. D.",
    title = "{Massive Primordial Black Holes}",
    eprint = "1911.02382",
    archivePrefix = "arXiv",
    primaryClass = "astro-ph.CO",
    doi = "10.22323/1.362.0013",
    journal = "PoS",
    volume = "MULTIF2019",
    pages = "013",
    year = "2020"
}

@article{Davoudiasl:2021ijv,
    author = "Davoudiasl, Hooman and Denton, Peter B. and Gehrlein, Julia",
    title = "{Supermassive Black Holes, Ultralight Dark Matter, and Gravitational Waves from a First Order Phase Transition}",
    eprint = "2109.01678",
    archivePrefix = "arXiv",
    primaryClass = "astro-ph.CO",
    doi = "10.1103/PhysRevLett.128.081101",
    journal = "Phys. Rev. Lett.",
    volume = "128",
    number = "8",
    pages = "081101",
    year = "2022"
}

@article{Carr:2021bzv,
    author = "Carr, Bernard and Kuhnel, Florian",
    title = "{Primordial black holes as dark matter candidates}",
    eprint = "2110.02821",
    archivePrefix = "arXiv",
    primaryClass = "astro-ph.CO",
    doi = "10.21468/SciPostPhysLectNotes.48",
    journal = "SciPost Phys. Lect. Notes",
    volume = "48",
    pages = "1",
    year = "2022"
}

@article{Belotsky:2017puc,
    author = "Belotsky, K. M. and Grobov, A. V. and Rubin, S. G.",
    title = "{Local heating of the universe by the Higgs field}",
    doi = "10.1142/S0218271818410031",
    journal = "Int. J. Mod. Phys. D",
    volume = "27",
    number = "06",
    pages = "1841003",
    year = "2017"
}

@article{Grachev:2010tn,
    author = "Grachev, S. I. and Dubrovich, V. K.",
    title = "{Propagation of the burst of radiation in expanding and recombining Universe: Thomson scattering}",
    eprint = "1010.4455",
    archivePrefix = "arXiv",
    primaryClass = "astro-ph.CO",
    doi = "10.1134/S1063773711040013",
    journal = "Astron. Lett.",
    volume = "37",
    pages = "293",
    year = "2011"
}

@article{Belotsky:2017txw,
    author = "Belotsky, K. M. and Golikova, Yu. A. and Rubin, S. G.",
    title = "{Local heating of matter in the early universe owing to the interaction of the Higgs field with a scalar field}",
    doi = "10.1134/S1063778817040056",
    journal = "Phys. Atom. Nucl.",
    volume = "80",
    number = "4",
    pages = "718--720",
    year = "2017"
}

\appendix
\section{Reaction rates and cross-sections}
\label{reac_app}
Here we 
calculate the thermonuclear reaction rates. 
Maxwell–Boltzmann distributions are assumed for 
interacting nuclei at thermodynamic equilibrium, therefore it follows that the relative velocities between the two species of nuclei will also be Maxwellian in nature \cite{clayton1983principles}. We may write for the Maxwell–Boltzmann distribution

\begin{equation}
P(v)dv=\Big(\frac{m_{12}}{2\pi T}\Big)^{3/2} \  e^{-m_{12}v^2/(2T)} \ 4\pi v^2 dv.
\end{equation}
where $m_{12} = m_{1}m_{2}/(m_{1}+m_{2})$ is the reduced mass (Boltzmann constant is assumed to be 1). With $E = m_{12}v^2/2$ and $dE/dv= m_{12}v$, we  may write the velocity distribution as an energy distribution,
\begin{align}
P(v)dv&=P(E)dE=\Big(\frac{m_{12}}{2\pi T}\Big)^{3/2} \  e^{-E/T} \ 4\pi \frac{2E}{m_{12}} \frac{dE}{m_{12}} \sqrt{\frac{m_{12}}{2E}}\nonumber\\
&=\frac{2}{\sqrt{\pi}} \frac{1}{(T)^{3/2}} \sqrt{E} \ e^{-E/T} dE
\end{align}
For the reaction rate we obtain \cite{iliadis2015nuclear}
\begin{align}\label{sigmv12}
\langle \sigma v\rangle_{12}=\int_{0}^{\infty}v \ \sigma(v) \ P(v)dv
=\Big(\frac{8}{\pi m_{12}}\Big)^{1/2} \ \frac{1}{(T)^{3/2}} \int_{0}^{\infty}E \ \sigma(E) \ e^{-E/T}dE.
\end{align}
The rate of the reaction is significantly dependent on the cross section $\sigma$, which varies for each nuclear reaction.\\


The reaction rates can be calculated using either numerical integration or analytical formulas used in this section. At this stage, we define the astrophysical S-factor \cite{angulo1999compilation}, $S(E)$, as
\begin{equation}
\sigma(E) \equiv \frac{1}{E} \ e^{-2\pi \eta}  \ S(E)
\end{equation}

Remember that the Gamow factor $e^{-2\pi \eta}$ is just a rough approximation for the s-wave transmission probability for energies considerably below the Coulomb barrier height. We write for the reaction rate using the S-factor definition.
\begin{multline}\label{SE}
n_1n_2\big\langle \sigma v\big\rangle_{12}=\Big(\frac{8}{\pi m_{12}}\Big)^{1/2} \ \frac{n_1n_2}{(T)^{3/2}} \int_{0}^{\infty} \ \text{exp}\Big( -\frac{2\pi}{\hbar}\sqrt{\frac{m_{12}}{2E}}Z_1Z_2 e^2\Big)  \ S(E) \ e^{-E/T}dE
\end{multline}
where $Z_i$ is the charges of target and projectile. 
The energy dependency of the integrand is remarkable. The term $e^{-E/T}$, derived from the Maxwell Boltzmann distribution, approaches zero for high energy, whereas the term $e^{-1/\sqrt{E}}$, derived from the Gamow factor, approaches zero for low energies. The most significant contribution to the integral will come from energies where the product of both terms is near its maximum.\\

Correction is required here since the S-factor for many reactions is not constant but changes with energy. In most situations, just expanding the experimental or theoretical S-factor into a Taylor series around $E=0$ is acceptable \cite{fowler1967thermonuclear,lang2013astrophysical}.

\begin{equation}
S(E) \approx S(0) + S^{'}(0)E + \frac{1}{2} S^{''}(0) E^2
\end{equation}
where the primes are derivatives with regard to $E$. Substituting this expansion into Eq.\eqref{SE} results in a sum of integrals, each of which may be extended into powers of $1/\tau$ ($\tau \equiv E_0/(T)$). Table 1 shows the values of the astrophysical S-factor, $S(E)$, for three reactions in proton-proton chains (ppI chain), that will be investigated in the next sections of this paper. 
\begin{table}[h]
\centering
\caption{\label{cross}Best-estimate low-energy nuclear reaction cross section factors \cite{angulo1999compilation,fowler1967thermonuclear}}
\begin{tabular}{||c|c|c|c||}
\hline
Reaction &  $S(0)$ MeV b& $S'(0)$ b & $S''(0)$ MeV$^{-1}$ b\\
\hline
p(p,e$^+\nu$)d & 3.94 $\times$ 10$^{-25}$ & 4.61$\times$ 10$^{-24}$ & 2.96$\times$ 10$^{-23}$ \\
\hline
p(n,$\gamma$)d & 7.30 $\times$ 10$^{-20}$ & -1.89$\times$ 10$^{-19}$ & 2.42$\times$ 10$^{-19}$ \\
\hline
d(p,$\gamma$)$^3$He & 0.20 $\times$ 10$^{-6}$ & 5.60 $\times$ 10$^{-6}$ & 3.10 $\times$ 10$^{-6}$ \\
\hline
$^3$He($^3$He,2p)$\alpha$ & 5.18 & - 2.22 & 0.80 \\
\hline
\hline
\end{tabular}
\end{table}

For reaction with participation of $e^{\pm}$ the following approximate formulas are used
\begin{eqnarray}
\sigma_{en}=\sigma_{ee}=\sigma_w,\;\;\;\;\;\;\;\;\;\;\;\;\;\;\;\;\;\;\;\;\;\;\;\;\;\sigma_{ep}=\sigma_w\, \exp\left(-\frac{Q}{T}\right), \label{EE}\\
\sigma_w \sim G_{\rm F}^2 T^2,\;\;\;\;\;\;\;\;\;\;\;\; Q=m_n-(m_e+m_p)=0.77 \; \;\text{MeV}.\label{SW}
\label{sigma}
\end{eqnarray}
They effectively take into account the threshold effect in respective reaction, $G_{\rm F}=1.166\times 10^{-5}\,\rm{ GeV}^{-2}$ is the Fermi constant. Such an estimation of cross section has accuracy factor $3$ of in the range where it is relevant. The reaction rates of the light elements are calculated by Eq.\eqref{SE}.\\

\begin{figure}[h]
    \centering
    \includegraphics[width=0.9\textwidth]{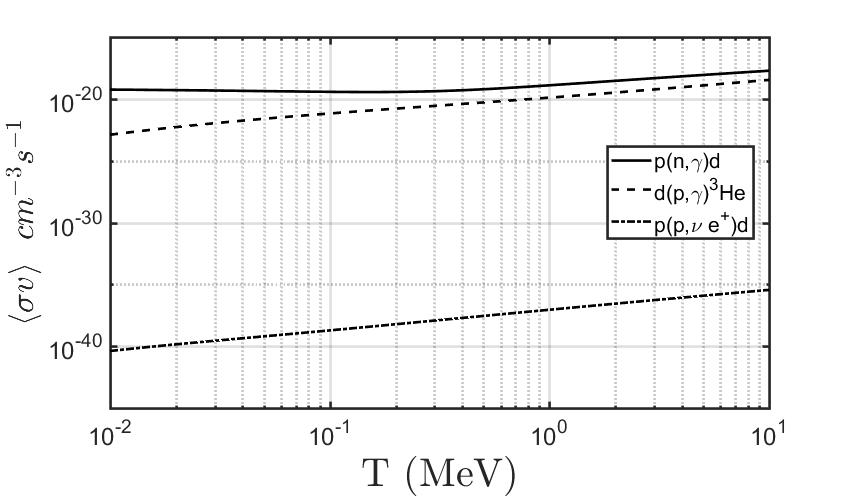}
    \caption{Reaction rate versus temperature.}
    \label{ST}
\end{figure}

Figure \ref{ST} shows reaction rates of p(n,$\gamma$)d, d(p,$\gamma$)$^3$He and   p(p,e$^+\nu$)d with temperature, which calculated by Eq.\eqref{SE}. 

\end{document}